\documentclass[useAMS,usenatbib,letterpaper]{mn2exxx}
\usepackage{times,psfig,url}

\title[Supernova remnant energetics and magnetars]
{
Supernova remnant energetics and magnetars: no evidence in favour
of millisecond proto-neutron stars
}

\author[J. Vink \& Lucien Kuiper]{Jacco Vink$^{1,2}$\thanks{E-mail: j.vink@astro.uu.nl}
 and Lucien Kuiper$^{2}$\\
$^{1}$Astronomical Institute, Utrecht University,
           PO Box 80000, 3508 TA Utrecht, The Netherlands\\
$^{2}$SRON Netherlands Institute for Space Research, Sorbonnelaan 2, 
3584CA Utrecht, The Netherlands\\
}

\begin{document}

\date{}
\pagerange{\pageref{firstpage}--\pageref{lastpage}} \pubyear{2005}

\maketitle

\label{firstpage}

\newcommand{\rosat}{{\em ROSAT}}
\newcommand{\chandra}{{\em Chandra}}
\newcommand\asca{{\em ASCA}}
\newcommand\xmm{{\em XMM-Newton}}
\newcommand\sax{{\em BeppoSAX}}
\newcommand\rxte{{\em RXTE}}
\newcommand\msun{{$M_{\odot}$}}
\newcommand{\kms}{{km\,s$^{-1}$}}
\newcommand{\net}{{$n_{\rmn e}t$}}

\newcommand\spex{{SPEX}}
\newcommand\xspec{{XSPEC}}

\newcommand{\adspr}{{AdSpR}}
\newcommand{\apj}{{ApJ}}
\newcommand{\apjs}{{ApJS}}
\newcommand{\apjl}{{ApJ}}
\newcommand{\aj}{{AJ}}
\newcommand{\aap}{{A\&A}}
\newcommand{\aaps}{{A\&AS}}
\newcommand{\nat}{{Nat}}
\newcommand{\jetp}{{JETP}}
\newcommand{\mnras}{{MNRAS}}
\newcommand{\phrvl}{{PhRvL}}
\newcommand{\phrc}{{PhRvC}}
\newcommand{\prc}{{PhRvC}}
\newcommand{\araa}{{ARA\&A}}
\newcommand{\pasj}{{PASJ}}
\newcommand{\pasp}{{PASP}}
\newcommand{\npa}{{NuPhA}}
\newcommand{\iaucirc}{{IAU circ.}} 
\newcommand{\aplett}{{Astrophysical Letters}} 
\newcommand{\gca}{Geochimica et Cosmochimica Acta}

\begin{abstract}
It is generally accepted that Anomalous X-ray Pulsars (AXPs) and 
Soft Gamma-ray Repeaters (SGRs) are magnetars, i.e. neutron stars with
extremely high surface magnetic fields ($B > 10^{14}$~G). 
The origin of these high magnetic fields is uncertain,
but  a popular hypothesis is that magnetars are born with an initial spin 
period not much exceeding the convective overturn time ($\sim 3$~ms),
which results in a powerful  dynamo action, amplifying the seed 
magnetic field to $\ga 10^{15}$~G.
Part of this rotation energy is then expected to
power the supernova through rapid
magnetic  braking. It is therefore possible that magnetars 
creation is accompanied by supernovae that are an order of magnitude
more energetic than normal supernovae, provided their initial 
spin period is $\sim 1$~ms.
However, we list here evidence that 
the explosion energies of these supernova remnants
associated with AXPs and SGRs -- Kes 73 (AXP 1E 1841-045), 
CTB 109 (AXP 1E2259+586) and N49 (SGR 0526-66) -- are close
to the canonical supernova explosion energy of $10^{51}$~erg,
suggesting an initial spin period of $P_0 \ga 5$~ms. 

We therefore do not find evidence that magnetars are formed
from rapidly rotating proto-neutron stars, allowing for the
possibility that they descend from stellar progenitor
with high magnetic field cores, and we discuss the merits of both formation 
scenarios.

In an appendix we describe the analysis of XMM-Newton
observations of Kes 73 and N49 
used to derive the explosion energies for these remnants.

\end{abstract}

\section{Introduction}
The notion that neutron stars exist with surface  magnetic fields as high as 
$10^{14}-10^{15}$~G has become generally accepted over the last decade.
The most spectacular manifestations of these so-called magnetars are
arguably the Soft Gamma-ray Repeaters (SGRs), which have 
received ample attention
after the giant flare of SGR\,1806-20 on December 27, 2004, 
which had an energy of $\sim 10^{46}$~erg \citep[e.g.][]{hurley05}. 
Related to the class of SGRs are the Anomalous X-ray Pulsars (AXPs), 
which are less prone to flare, although occasional flares have been observed
\citep{gavriil02,kaspi03}. 
Like SGRs they are radio quiet X-ray pulsars with spin 
periods in the range of 5-12 s.
The timing properties of both SGRs and AXPs suggest that they are isolated 
neutron stars, whereas their spin down rate suggest that they have 
magnetar-like dipole magnetic fields of $B_{\rm dip}\sim 10^{14}-10^{15}$~G 
\citep[see][for a review]{woods04}. Other arguments 
in favour of the magnetar hypothesis can be 
found in \citet{thompson95}.

The apparent dichotomy between the radio quiet magnetars, on the one hand,
and ordinary young radio pulsars, with $B\sim 10^{12}-10^{13}$~G, 
on the other hand, is likely to originate from distinct properties of the
progenitor stars.
One idea is that the high magnetic field of magnetars simply
reflects the high magnetic field of their progenitor stars. Magnetic flux
conservation \citep{woltjer64} implies that magnetars must then be the
stellar remnants of stars with internal magnetic fields of 
$B \ga 10^4$~G,
whereas radio pulsars must be the end products of stars with  
$B \sim 10^3$~G.
Based on a study of magnetic white dwarfs, \citet{ferrario06} argue that
there is a wide spread in white dwarf progenitor magnetic fields, which, 
when extrapolated to the more massive 
progenitors of neutron stars can explain the existence of magnetars.

\begin{table*}
\begin{minipage}{0.92\textwidth}
\caption{The explosion energies and ages of the supernova remnants 
from X-ray spectral analysis. 
}
\label{tab-energies}
\begin{tabular}{l c c c c c c c l}\hline\hline\noalign{\smallskip}
SNR/Pulsar & Distance & radius & E &  $n_{\rmn H}$ & Mass & SNR Age & Pulsar Age & References\\
    &kpc   & pc  & ($10^{51}$~erg) & cm$^{-3}$ & \msun & $10^3$yr & $10^3$ yr\\
\noalign{\smallskip} \hline\noalign{\smallskip}
Kes 73/1E1841-045 & 
$7.0$& 4.3 & $0.5\pm0.3$ & $2.9\pm0.4$ & $29\pm4$ & $1.3\pm0.2$&  4.3 & This work \\
CTB109/1E2259+586 &
$3.0$& 10 & $0.7\pm0.3$ & $0.16\pm0.02$ & $97\pm23$ & $8.8\pm0.9$&  220 & \citet{sasaki04}\\
N49/SGR 0526-66 & 
$50$& 9.3 & $1.3\pm0.3$ & $2.8\pm0.1$ & $320\pm50$ & $6.3\pm$1.0&  1.9 & This work\\
\noalign{\smallskip}\hline
\end{tabular}\\
Note that \citet{sasaki04} find an energy of $(1.9\pm0.7)\times 10^{51}$~erg 
for CTB109, if they assume
incomplete temperature equilibration at the shock front. Derived energies
scale with distance $d$ as $d^{2.5}$. 
Distances and pulsar ages ($\tau = \frac{1}{2}P/\dot{P}$) are taken from 
\citet{woods04}.
\end{minipage}
\end{table*}

A currently more popular idea for the origin of magnetars is
that they are formed from proto-neutron stars with periods
in the range  $P_i\sim 0.6$~ms (the break-up limit) to 3~ms.
Convection and differential rotation in such rapidly rotating stars would
give rise to an efficient $\alpha-\Omega$-dynamo, 
resulting in magnetic field amplification on a time scale of 
$\la 10$~s \citep{duncan92,thompson93,duncan96}. 
In contrast, radio pulsars presumably
start their life with initial spin periods of  $\ga 10$~ms or longer.
As initially pointed out by \citet{duncan92},
the idea that magnetars form from proton-neutron
stars with $P\sim1$~ms implies that supernova explosions that create magnetars
are an order of magnitude more energetic than ordinary core collapse
supernovae.
The reason is that a neutron star spinning with $P = 1$~ms 
($\Omega =6.3\times10^3$~s$^{-1}$)  has a rotational energy of 
$E_{rot} = \frac{1}{2} I \Omega^2\approx 3\times10^{52}
\bigl(\frac{P}{1\,\rm  ms}\bigr)^{-2}$~erg, 
for a moment of inertia of $I \approx 1.4\times10^{45}$~g\,cm$^2$ 
\citep{lattimer01}. 
Due to the high magnetic field, the magnetar rapidly spins down during 
the explosion, thereby powering the supernova. 
An upper limit for the time scale in which most of the rotational energy
is lost by a neutron star with $B=10^{15}$~G is obtained by assuming 
magnetic braking in vacuum, $\tau \approx 160$~s.
More detailed modeling by \citet{thompson04} indicates that
the time scale for spin down of the proto-neutron star may be as short as
10-100~s.

Given the implication of the millisecond proto-neutron star 
hypothesis that magnetars should be surrounded by {\em hyper}nova remnants,
it is surprising that this idea has not directly 
been tested on supernova remnants (SNRs)
associated with AXPs and SGRs, although \citet{allen04} did calculate
the effect of the hypernova/magnetar connection on the evolution
of the SNR, and applied this to known magnetar/SNR connections.
Here we 
review what is known about the explosion energies of the supernova remnants
Kes 73, N49 and CTB 109 associated with respectively the magnetar candidates
{AXP 1E1841-045}, {SGR 0526-66}, and {AXP 1E2259+586}.
For CTB 109 and N49 explosion energy estimates have been given in the 
literature, but without discussing the possible implications for the
magnetar formation scenario. As far as we know, no explosion energy estimates
exists for Kes 73, we therefore present our own energy estimate based on
archival \xmm\ observations. In addition we present an analysis of 
the explosion energy of N49, also based on unpublished \xmm\ data.

\begin{figure*}
\centerline{
\psfig{figure=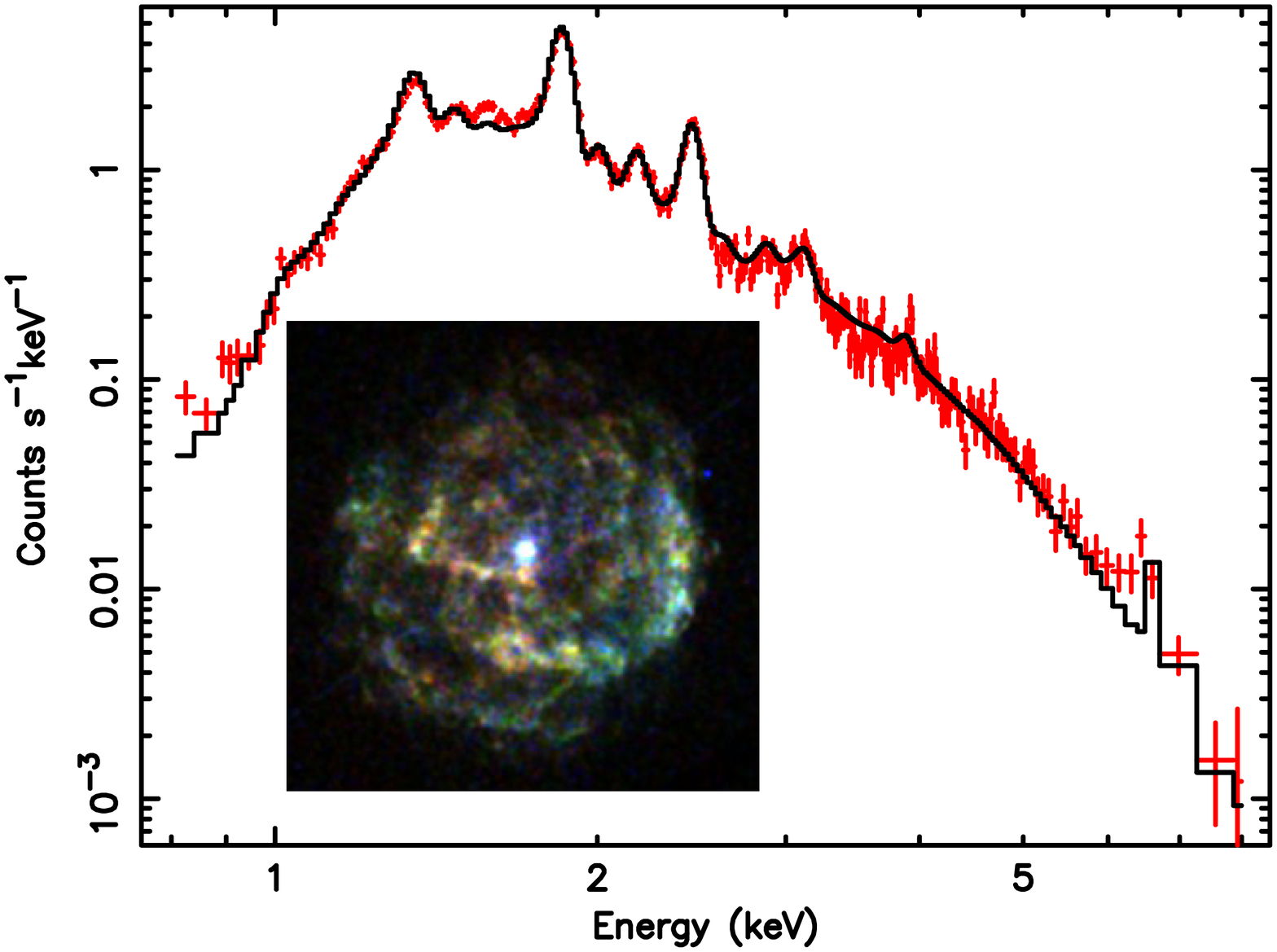,width=0.45\textwidth}
\hskip 0.05\textwidth
\psfig{figure=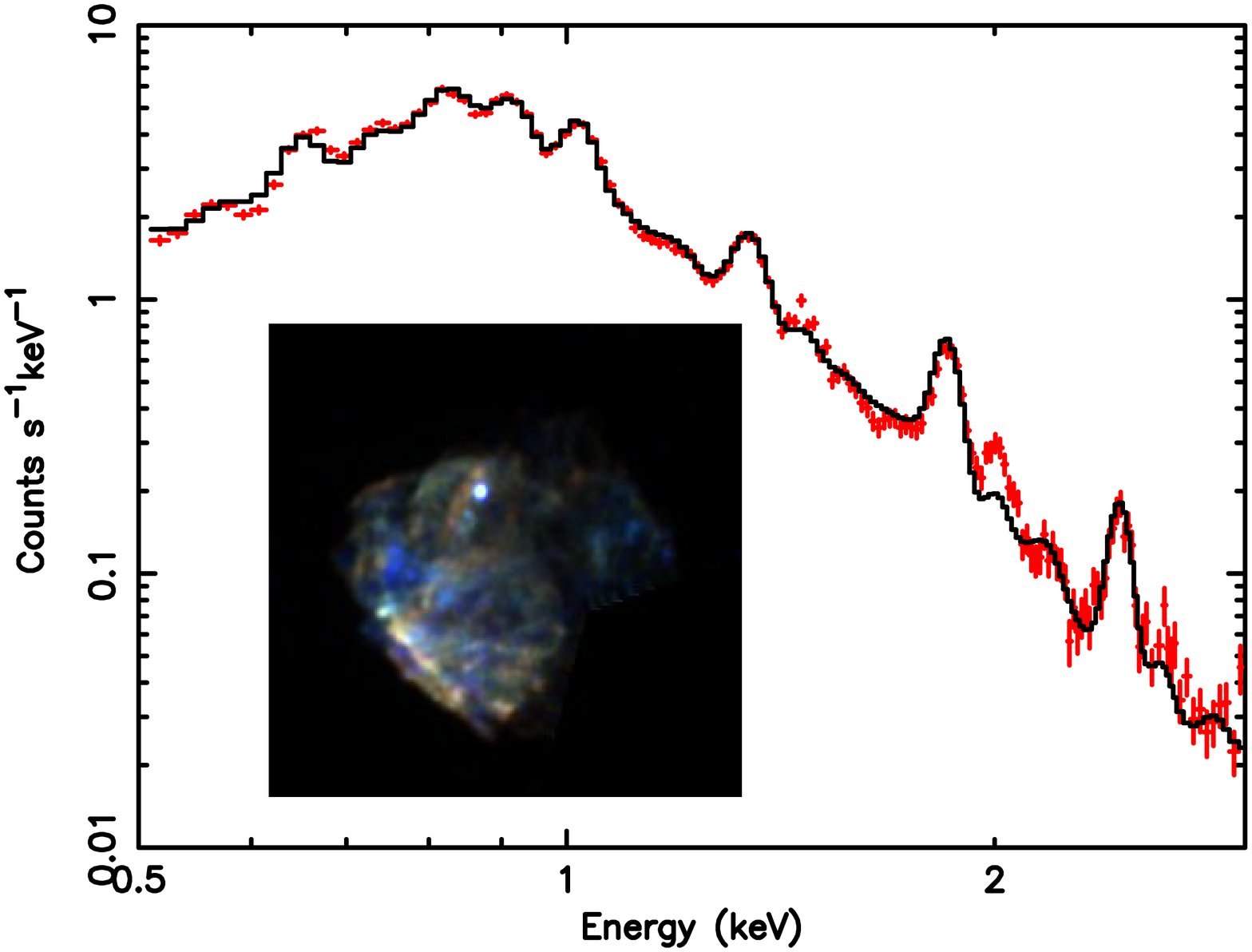,width=0.45\textwidth}
}
\caption{\xmm\ MOS12 spectra of Kes 73 (left) and N 49 (right). The solid lines
represent the best fit Sedov models. See appendix for details. 
Chandra ACIS images (taken from the archive) are shown as insets.
The RGB color coding for Kes 73 corresponds to the
energy ranges: 0.6-1.5 keV, 1.5-2.8 keV, 
2.8-5 keV. The image measures 5.1\arcmin$\times$5.1\arcmin. 
The N49 images consists of a mosaic, and the RGB colors correspond
to 0.5-0.75 keV,
0.75-1.2 keV and 1.2-3 keV. The image measures 1.9\arcmin$\times$1.9\arcmin,
but due to the small window mode of the observations the southwestern
corner of N49 is missing. AXP 1E1841-045 is the bright unresolved source in the
center of Kes 73; SGR 0526-66 is the unresolved source near the northern
edge of N49.
For both images a linear brightness scaling
was used, but saturating the unresolved sources in order to bring out
the SNR emission. 
}
\label{fig-spectra}
\end{figure*}

\section{The explosion energies of supernova remnants associated 
with magnetars}
\citet{gaensler04} reviewed the association of AXPs and SGRs with SNRs, taking
into account the probability of random spatial coincidence
given the extent of the SNR
and the location of the AXP or SGR within the SNR. He concludes that
magnetar candidates likely to be associated with SNRs (i.e. have
a chance alignment probability $<$ 1\%) are: 
{AXP 1E1841-045} with Kes 73 (G27.4+0.0), 
{AXP 1E2259+586} with CTB 109 (G109.1-1.0), 
SGR 0526-66 with N49, and the candidate {AXP AX J1845-045} with G29.6+0.1.

G29.6+0.1, a shell-type SNR, was detected in archival radio data
\citep{gaensler99}  following the discovery of AX J1845-045 \citep{gotthelf98}.
Unfortunately, the X-ray emission from the SNR is too weak \citep{vasisht00}
and the distance too poorly constrained \citep{gaensler99} 
to derive any meaningful energy estimate from the X-ray
spectrum of the SNR. This leaves us with Kes 73, N49 and CTB 109
to test whether SNRs with magnetars are an order of magnitude
more energetic than regular SNRs.

One can determine the energy of SNRs using the well known Sedov
evolution of an adiabatic point explosion 
in a uniform medium \citep{truelove99}.
The shock radius, $r_s$, and velocity, $v_s$, of such an explosion evolves as
\begin{equation}
r_s^5 = \frac{2.026 E t^2}{\rho_0},\label{sedov}
\end{equation}
\begin{equation}
v_s = \frac{2 r_s}{5t},\label{sedov2}
\end{equation}
with $\rho_0$ the interstellar medium density, $E$ the explosion energy,
and $t$ the age of the SNR.
The shock velocity can be determined from X-ray spectra, because
for a monatomic gas and high Mach number shocks the following
relation holds between post-shock plasma temperature and shock velocity:
\begin{equation}
kT = \frac{3}{16}0.63 m_{\rmn p} v_s^2, \label{temp}
\end{equation}
with $m_{\rmn p}$ the proton mass. The factor 0.63 gives the average
particle mass in units of the proton mass. It is smaller
than the proton mass, because it includes electrons.
So from measuring the temperature one can estimate $v_s$, which, combined
with the measured radius, gives the age of the SNR (Eq.~\ref{sedov} \& 
\ref{sedov2}).
The interstellar medium density can be estimated from the
X-ray emission measure (EM $= \int n_{\rmn e}n_{\rmn H}dV$).
Having obtained a density estimate, the explosion energy can be calculated
using Eq.~\ref{sedov}.

We list the results of our analysis of \xmm\ data 
in Table~\ref{tab-energies}, together with the parameters for CTB 109 reported
by \citet{sasaki04}. 
It is clear that these SNRs with magnetars do not
have an order of magnitude higher explosion 
energies than the canonical supernova explosion energy of $10^{51}$~erg,
suggesting an initial period $P \ga 5$~ms.
More details about our analysis
data can be found in the appendix, which 
also includes a short discussion on the effects of non-equilibration of 
temperatures
on the derived parameters, and on the applicability of the Sedov
model for Kes 73. Figure \ref{fig-spectra} shows the \xmm\ spectra of Kes 73
and N49 with best fit spectral models.

For all three SNRs the distances are well determined. N49
is located in the Large Magellanic Cloud, Kes 73 has a reliable
distance estimate based on HI absorption measurements 
\citep[$d=6-7.5$~kpc,][]{sanbonmatsu92}, and CTB 109 cannot be more
distant than Cas A, which is situated
at 2\degr\ from CTB 109, has a higher absorption column,
and is at a distance of 3.4~kpc \citep{reed95}.
Both TB109 and Cas~A 
are likely to be associated with the Perseus arm at $\sim 3$~kpc.

As an aside, Cas A contains a point source that is suspected to be a magnetar
\citep[see][for a discussion]{pavlov04}, 
a suspicion that has gained more credibility with the
recent discovery of an infra-red light echo from a putative
SGR-like giant flare \citep{krause05}. However, Cas A appears to have an
explosion energy of $\sim 2\times10^{51}$~erg, which is quite accurately known,
as the expansion has been directly measured \citep[][for a review]{vink04a}.

We like to conclude this section by pointing out that, although there are some
uncertainties associated with determining the explosion energies using a Sedov
analysis (gradients in the medium in which
the SNR evolves, conversion of part of the energy to cosmic rays), 
these uncertainties apply equally to all other SNRs for which energies 
have been determined. Nevertheless, other SNRs appear to have similar
explosion energies \citep{hughes98}.
In other words, 
the most remarkable feature of SNRs with magnetars is that they
are unremarkable, i.e. they are indistinguishable
from normal SNRs.

\section{Discussion}

This study shows that {\em in general} magnetar formation is not
accompanied by a hypernova explosion.
This does
not necessarily preclude the idea that observed extra-galactic hypernovae,
such as  SN1998bw, are not by the formation of a rapidly
spinning magnetar followed by rapid magnetic braking
\citep{nakamura98,thompson04}, but it poses a challenge
to the hypothesis that magnetars are formed as a result of magnetic field
amplification in a rapidly spinning ($P \sim 1$~ms) proto-neutron star.

Note that when the magnetar theory of SGRs and AXPs was first proposed, 
the existence of magnetars was hardly an established fact 
\citep{duncan92,thompson93}.
Today their existence is only doubted by a few, especially after the
giant flare of SGR\,1806-20. In general, SGR flare luminosities
exceed the Eddington luminosity, making it hard to find any other
source of energy but the internal magnetic field. The energy of the giant
flare of SGR\,1806-20 was $\sim 10^{46}$~erg \citep{hurley05}. If a neutron
star has several of those flares this energy can only be provided with
an internal magnetic field of $\sim 10^{16}$~G \citep[e.g.][]{stella05}.
Moreover, the results presented do not rule out 
initial periods of $\sim 3$~ms,
the upper limit at which the $\alpha-\Omega$\ dynamo may operate
\citep{duncan96}.

The generation of such a high internal magnetic field poses challenges in 
itself. The magnetar-sized 
magnetic fields may either derive from a high magnetic field in the core
of the neutron star progenitor (the fossil field hypothesis), 
or may be generated in the proto-neutron star, in which case rapid rotation
and strong convective motions and/or differential rotation are needed.
It is indeed shown that proto-neutron stars are highly convective
\citep{buras06,fryer04}, although \citet{fryer04} express some doubt
whether it is sufficient to generate magnetic fields in excess of
$10^{14}$~G. The rapid rotation needed for amplifying magnetic fields in
proto-neutron stars may reflect the high angular momentum of the
progenitor's core. However, the proto-neutron star may also wind up
due to asymmetric accretion during collapse, in which case it is expected
that the accretion results in both additional spin and kick velocity
\citep{spruit98,thompson00}.

For the fossil field hypothesis, the solution to 
the question  what the origin of the high magnetic field is, 
creates itself  a new problem,namely, what is the origin of the high magnetic 
field of the progenitor star?
It could be that the magnetic field
is again the result of the amplification of a seed magnetic field due
to rotation and convection, but in this case in the core of the progenitor.
However, as noted by \citet{thompson93}, there is less convective
energy available during any stage of stellar evolution, than during the
proto-neutron star phase. Another option is what one might call the
strong fossil field hypothesis: the magnetic field in the star reflects
the magnetic field of the cloud from which the star was formed.

Ironically, stellar evolution models indicate 
that progenitors with a high magnetic field in the core end up producing
more slowly rotating pulsars \citep{heger05}.
This is a result of the more effective coupling
of angular momentum of the core to the star's outer radius, where
angular momentum is lost due to stellar winds.

The results presented here are in agreement with the fossil field hypothesis.
Recently, \citet{ferrario06} showed that the magnetic flux distribution
of magnetic white dwarfs  and neutron stars are similar. Translated to neutron
star radii, the highest magnetic field measured in white dwarfs correspond to
$10^{14}$-$10^{15}$~G. However, this falls still short of the $10^{16}$~G
needed to power flares from SGRs such as SGR 1806-20.
Possibly this could be accounted for by additional magnetic field amplification
during the late stages of stellar evolution of stars that end in a 
core collapse
\citep[c.f. the core magnetic field evolution of the magnetic star of][]{
heger05}.
On the other hand, the energetics of the SNRs with magnetars could still
be reconciled with magnetic field amplification in a rapidly rotating 
proto-neutron star, provided the angular momentum is not lost by
magnetic braking, in which case it powers the supernova, but due
to gravitational radiation or due to emission in a jet 
\citep[see also][]{thompson93}. Such a loss mechanism should
work on a time scale shorter than the magnetic braking time scale, i.e.
less than a few hundred seconds.

\citet{stella05}
argued recently that if {\it internal} magnetic fields exceed
$5\times10^{16}$~G neutron stars deformation is sufficiently strong
to cause gravitational waves that radiate away most of  the rotational energy.
Note, however, that this is only true if rotation losses due 
gravitational waves exceed losses due to magnetic braking.
This requires $B_i>5\times 10^{16}$~G, 
for which no conclusive evidence exists yet, whereas
the external dipole field has to be low, 
$B_d<5\times10^{14}$~G. Moreover, this estimate is based on magnetic
braking in vacuum, whereas during the first few seconds
magnetic braking is likely to be enhanced 
due to the presence of hot plasma surrounding the proto-neutron star 
\citep{thompson04}.

It is possible that in the early phase of a
magnetar the dipole field is indeed low. On the other hand, the observed
dipole fields are lower than is required to slow down to the spin periods
of AXPs and SGRs. A case in point is 1E1841-045/Kes73, for which we
determined an age of $\sim1300$~yr. 
To bring it from a period of few millisecond to its current
period (11.8 s) a dipole field is required of $1.6\times10^{15}$~G
for a braking index of $n=2.5$. Gravitational waves cannot explain
this, since only for very short periods
do gravitational wave losses
dominates over magnetic braking, because rotation loss
scales with $\Omega^5$ for gravitational waves and with  $\Omega^3$ for
magnetic braking. 

Another consequence is that as soon
as magnetic braking dominates one expects the creation of a strong
a pulsar wind nebula.
Intriguingly, none of the magnetars studied here
shows any evidence for a pulsar wind component. The radio map of Kes 73 shows, 
in fact, that the
AXP is located in a hole in the SNR \citep{kriss85}. 
In X-rays we do not expect to see a pulsar wind nebula, 
as the ultra-relativistic electrons responsible for X-ray synchrotron emission
will have lost most of their energy.
Note that the spin-down of a magnetar like  1E1841-045 is so rapid
that the pulsar wind nebula must evolve during the early, free expansion
phase of the SNR. Since most of the ejecta has not been shocked it is possible
that the relativistic particles freely escape instead of being confined
by the hot plasma of the SNR. However, it would be of great importance 
to investigate the
formation of pulsar wind nebulae from magnetars in more detail, as it may
provide additional means of determining whether  AXP/SGRs
started with periods in the milli-second range.

Finally, another way of losing angular momentum is by means of a
jet during the explosion. Radio and X-ray maps of Kes 73 and N49 do
not show evidence that a jet may have been present in the past. Kes 73
is in fact remarkably circular symmetric (Fig.~\ref{fig-spectra}).
Interestingly, however, there is considerable evidence for a jet in
the SNR Cas A 
\citep{vink04a,hwang04}, and there is
indeed evidence, although not conclusive, that the point source
in Cas A is a magnetar \citep{krause05}.

\section*{Acknowledgments}
JV thanks Andrew MacFadyen, Nanda Rea, Robert Duncan and the referee,
Chris Thompson, for helpful
discussions and suggestions.


\appendix
\section{The spectral analysis of XMM-Newton data of Kes 73 and N49}

\begin{table}
\centering
\caption{Best fit spectral models to \xmm/MOS1 data.}
\label{tab-sedov}
{\footnotesize
\begin{tabular}{lccc}\hline\hline
                   &\multicolumn{2}{c}{{\normalsize Kes 73}} & {\normalsize N49}\\
Model              & 2 NEI & Sedov & Sedov \\
\hline\noalign{\smallskip}

EM  ($10^{12}$cm$^{-5}$)          & $26.9\pm3.1$   & $16.4\pm1.5$     & $5.4\pm0.5$\\
$kT_{\rm e}$ (keV)                & $0.63\pm0.06$  & $0.72\pm0.3$ &$0.415\pm0.005$\\
\net\  ($10^{11}$cm$^{-3}$s)  & $3.1\pm0.8$    & $4.6\pm0.2$  & $28.0\pm2.6$\\

EM$_2$  ($10^{12}$cm$^{-5}$)      & $2.0\pm0.5$&\\
$kT_{2\rmn e}$ (keV)              & $2.26\pm0.35$  &\\
\net$_2$\  ($10^{11}$cm$^{-3}$s)  & $0.47\pm0.10$  & \\
O                                 & (1)     & (1)  & $0.25\pm0.02$\\
Ne                                & (1)     & (1)  & $0.41\pm0.02$\\
Mg                                & $1.13\pm0.13$  & $1.95\pm0.15$& $0.27\pm0.01$ \\
Si                                & $1.09\pm0.07$  & $1.7\pm0.1$  & $0.45\pm0.01$\\
S                                 & $1.12\pm0.06$  & $1.6\pm0.1$  & $0.99\pm0.06$\\
Ar                                & $1.02\pm0.18$  & $1.1\pm0.1$  & -\\
Ca                                & $1.84\pm0.42$  & $0.77\pm0.33$& -\\ 
Fe                                & $0.42\pm0.15$  & $4.3\pm0.7$  & $0.19\pm0.01$\\
\smallskip
$N_{\rmn H}$ ($10^{21}$cm$^{-2}$)  & $27.3\pm0.6$& $31.2\pm0.6$ & $2.2\pm0.1$\\
\smallskip
Fit range (keV)    &  0.8--8.0  & 0.8--8.0 & 0.5--3.0\\
C-statistic/d.o.f. &186/110 & 956/468& 574/153\\

\noalign{\smallskip}\hline
\end{tabular}\\
}
\medskip
\flushleft
NOTE - Errors correspond to 68\% confidence ranges ($\Delta C = 1.0$).
EM refers to the Emission Measure ($\int n_{\rmn e} n_{\rmn H}dV/(4\pi d^2)$).
Abundances are given with respect to the solar abundances of \citet{anders89}.
For Kes 73 the abundance of oxygen and neon were fixed to solar values,
because the strong absorption did not allow for an accurate determination.\\
\end{table}
For determining the energies of Kes 73 and N49 
we used archival \xmm\ data observations
made on October 5 and 7 2002 (Kes 73,ObsIDs 0013340101 \& 0013340201)
and of N49 made on July 7 2000 and April 8 2001 (N49,
ObsIDs 0111130301  \& 0113000201).
We only  used data from the MOS1 and MOS2 instruments, because they
offer the best energy resolution \citep{turner01}.
We processed the data in a standard manner employing the \xmm\ software 
SAS 6.5.0,
excluding time intervals with excessive background flares. The source
spectra were extracted from circular regions encompassing the source,
excluding small circular regions containing the AXP or SGR.
Background spectra were extracted from annular regions around the
SNRs. For our final spectral analysis we added all the MOS1 and MOS2 spectra 
of  each individual SNR together for statistical reasons,
using exposure weighted merged response files (RMFs) and 
ancillary response files (ARFs), a procedure that makes for a nicer 
representation of the spectra and gives little loss of accuracy, 
as the two MOS detectors are similar.
The total effective exposure per detector is 9.8 ks for Kes 73 and 39.6 ks for 
N49.

For the spectral analysis two spectral analysis packages were employed,
\spex\ \citep{kaastra00} and \xspec\ \citep{xspec}.
\spex\ has the advantage that it includes more lines associated with inner 
shell ionization, which is important for non-equilibration ionization (NEI) 
plasmas. 
This is in particular relevant for Kes 73, which is more out of ionization 
equilibrium 
than N49\footnote{
For a recent detailed X-ray study of this SNR see \citet{park03}} 
(Table~\ref{tab-sedov}). \xspec\ contains a Sedov model 
\citep[vsedov,][]{borkowski01b} that also
includes non-equilibration of electron and ion temperatures, an effect that
is known to be important for in particular SNRs with a short ionization age 
(\net) and high shock velocity \citep{rakowski05}.
We used \spex\ for a more heuristic approach in which we fit the spectra with 
two non-equilibration ionization models with two different temperatures and 
ionization ages, but with elemental abundances of the two components coupled.

From the results listed in Table~\ref{tab-sedov} one can calculate
the pre-shock density using the relation for SNRs in the Sedov phase
$\int n_{\rmn e} n_{\rmn H}dV = 0.129 \frac{4\pi}{3}r_s^3$. 
The shock velocity can be obtained from Eq.~\ref{temp}, and using the relation
between $v_s$, $r_s$ and $t$ (Eq.~\ref{sedov}) one can then infer the age of 
the SNR. From the age and the pre-shock density an estimate
of the explosion energy can be derived (Eq.~\ref{sedov}).
However, Eq.~\ref{temp} does not apply in case the plasma is not in full 
temperature equilibration. Although both remnants have lower shock velocities 
and much larger
ionization ages than a SNR like SN1006 for which this effect has in fact been
measured \citep[\net $= 2\times 10^9$~cm$^{-3}$s,][]{vink03b}, 
there may  still be some effect, in particular
for Kes 73. The plasma parameters of Kes 73 are in fact 
quite similar to those of CTB 109, for which \citet{sasaki04} argued that 
$T_{\rmn p }/T_{\rmn e} \le 2.5$. We find that for Kes 73 assuming 
non-equilibration gives a more consistent result. 
The reason is that the age of the
remnant can be estimated both from $kT$ and from the ionization age. 
From \net\ we derive 
$t = 1300\pm200$~yr, but using $kT$ we find from the values in the
second column of Table~\ref{sedov}
$2145\pm37$~yr for full equilibration, but $1357\pm23$~yr if 
$T_{\rmn p }/T_{\rmn e} = 2.5$. For N49 the ionization age and temperature give
consistent results assuming full equilibration 
(resp. $6.6\pm0.9$~kyr and $6.1\pm1.7$~kyr). 
This is to be expected, since the ionization age of N49 corresponds roughly to
the Coulomb equilibration time.

A note of caution concerning the interpretation of the X-ray spectrum 
of Kes 73:
As Table~\ref{tab-sedov} shows, there is 
considerable uncertainty about the abundance pattern of Kes 73, with the 2NEI
model suggesting abundances that are almost solar, whereas the Sedov model
suggest an enriched plasma. Other differences between the models in fact
go back to this, because fitting the Sedov model with the abundances
fixed at solar values we obtained a total emission measure and temperature 
closer to that of the 2NEI model. 
The matter is of some interest, as overabundances would
be an indication that Kes 73 is in an early evolutionary state that cannot be 
well
described by a Sedov model. Another possible indication for that is the low 
total mass of Kes 73 (25--33 \msun, for resp. the 2NEI and Sedov model),
much lower than N49 (320 \msun) or CTB 109 
\citep[$97\pm23$ \msun,][]{sasaki04}.
We consider the 2 NEI model to be more reliable 
in this case, since \spex\ incorporates
more line data relevant for NEI plasmas. In fact, Fig.~\ref{fig-spectra}
shows that the Fe-K complex around 6.6~keV and the Mg XII lines around 1.4 keV
are  not well fitted. The \spex\ 2NEI model does not have these problems.
Although, the fits to the spectra of Kes 73 and N49 are not perfect, the global
features of the spectra are well fitted (Fig.~\ref{fig-spectra}).

Nevertheless, we tried to constrain the explosion energy of Kes 73 further
by exploring the more generic SNR evolution models of \citet{truelove99}.
We tried to answer the question, which explosion energies are still consistent
with the densities and post-shock
temperatures found from our fit, 2--4~cm$^{-3}$ and $0.6-1.9$~keV,
given the shock radius of 4~pc and assuming that the 
ejecta masses lie in the range 0.5-25~\msun.
Considering ejecta density profiles with power indices 
of $n=0, n=7$, and $n=9$ 
\citep[see][]{truelove99}, it turns out that the observational data is only
consistent with energies in the range $(0.1-1.1)\times 10^{51}$~erg.

\end{document}